**Use of Non-concurrent Common Control in Master Protocols in Oncology Trials: Report of an American Statistical Association Biopharmaceutical Section Open Forum Discussion[+]**

Rajeshwari Sridhara[1*], Olga Marchenko[2], Qi Jiang[3], Richard Pazdur[1*], Martin Posch[4], Scott Berry[5], Marc Theoret[1*], Yuan Li Shen[1*], Thomas Gwise[1*], Lorenzo Hess[6*], Andrew Raven[7*], Khadija Rantell[8*], Kit Roes[9*], Richard Simon[10], Mary Redman[11], Yuan Ji[12], Cindy Lu[13]

[1]Oncology Center of Excellence US FDA, [2]Bayer, [3]Seagen, [4]Center for Medical Statistics, Informatics, and Intelligent Systems at the Medical University of Vienna, [5]Berry Consultants, [6]SMC, Switzerland, [7]HC, Canada, [8]MHRA, UK, [9]Radboud University Medical Center & Dutch Medicines Evaluation Board (CBG-MEB), [10]Simon Consulting, [11]Fred Hutch, [12]University of Chicago, [13]Biogen

*The opinions stated here are those of the authors and not necessarily of the Regulatory Agencies.

## Abstract

This article summarizes the discussions from the American Statistical Association (ASA) Biopharmaceutical (BIOP) Section Open Forum that took place on December 10, 2020 and was organized by the ASA BIOP Statistical Methods in Oncology Scientific Working Group, in coordination with the US FDA Oncology Center of Excellence. Diverse stakeholders including experts from international regulatory agencies, academicians, and representatives of the pharmaceutical industry engaged in a discussion on the use of non-concurrent control in Master

Protocols for oncology trials. While the use of non-concurrent control with the concurrent control may increase the power of detecting the therapeutic difference between a treatment and the control, the panelists had diverse opinion on the statistical approaches for modeling non-concurrent and concurrent controls. Some were more concerned about the temporality of the non-concurrent control and bias introduced by different confounders related to time, e.g., changes in standard of care, changes in patient population, changes in recruiting strategies, changes in assessment of endpoints. Nevertheless, in some situations such as when the recruitment is extremely challenging for a rare disease, the panelists concluded that the use of a non-concurrent control can be justified.

**Introduction**

The Biopharmaceutical Section (BIOP) of American Statistical Association (ASA) in coordination with the Oncology Center of Excellence (OCE), U.S. Food and Drug Administration (FDA) initiated a series of open-forum discussions on different aspects of statistical considerations for oncology clinical trials, aligning with the OCE's 'Project Signifi**CanT**' (Statistics in Cancer Trials). These open-forum discussions are designed to engage experts and diverse stakeholders who understand the unique aspects of oncology clinical trials. Issues discussed in these open-forum meetings can inform design and analysis of future oncology clinical trials.

The virtual open-forum discussion on use of non-concurrent common control in Master protocols in oncology trials was held on December 10, 2021 (Sridhara et al. 2021a). The panel consisted of diverse stakeholders and experts from international regulatory agencies, academia, and representatives of the pharmaceutical industry engaged in oncology product development. The discussions were moderated by the co-chairs of the Statistical Methods in Oncology Scientific Working Group of the BIOP, Qi Jiang, Ph.D. (Seagen), Olga Marchenko, Ph.D. (Bayer), and Rajeshwari Sridhara, Ph.D. (Lead for the Project **Signifi**CanT, Contractor at OCE FDA). The discussions focused on whether non-concurrent common control data can be used in temporal evaluation of multiple treatments under a Master Protocol oncology trial. The forum started with three formal presentations including an introduction, and frequentist and Bayesian approaches in using concurrent and non-concurrent common controls. The 17 panelists for the discussion included members of the BIOP Statistical Methods in Oncology Scientific Working Group representing pharmaceutical companies, representatives from international regulatory Agencies (US FDA, European Medicinal Agency (EMA), Health Canada (HC), Medicines and Healthcare products Regulatory Agency (MHRA, UK) and Swissmedic (SMC)), academicians and expert statistical consultants (see the agenda in Appendix). In addition, over 100 members attended the virtual meeting including representatives from other International Regulatory Agencies (e.g., from Japan, Australia, Singapore). In this report, we summarize the presentations and discussions briefly.

**Master Protocol**

Oncology drug development is impacting pharmaceutical industry, regulatory agencies and patients with the recent development of novel cancer drugs and growing interest in developing new drugs accounting for approximately 50% of FDA breakthrough designations.

The huge expansion of the number of new cancer therapies requires efficient trial designs that can address multiple clinical questions within a single study. Oncology is in a position to lead and promote the use of innovative designs that incorporate multiple treatments and multiple indications within a single Master Protocol. One such scenario is to consider a confirmatory trial with a Master Protocol after accelerated approval of drugs in a rare disease to enable efficient use of patient resources.

A Master Protocol establishes an infrastructure to potentially assess the therapeutic effects of one or multiple treatments for one or multiple disease conditions within a single trial (Woodcock and LaVange 2017; FDA 2018; Meyer et al. 2020). A Master Protocol often involves more complex statistical methodology and requires specialist expertise.

Master Protocols may use umbrella, basket, or platform designs. Platform trials allow temporal and staggered comparison of different treatment arms to a common/shared control arm (Fig.1). The data from the common control may be concurrent to a single treatment arm for the time period when the treatment is investigated in the trial or may be non-concurrent to other treatment arms that are investigated at different time periods. It is of interests to explore the use of concurrent and perhaps non-concurrent control data for statistical designs and analyses (Fig.2).

Figure 1. Master Protocol: Platform Trial

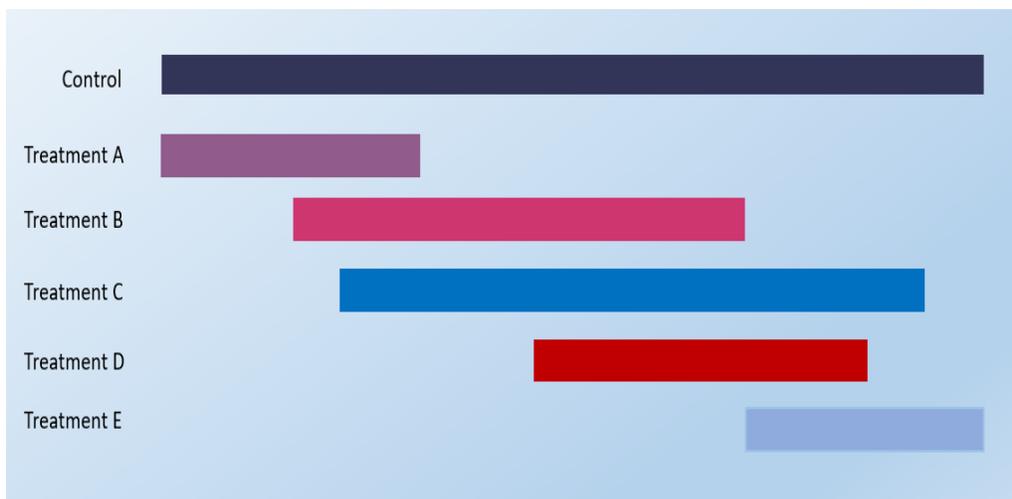

Figure 2. Non-Concurrent Controls (NCC) in a Platform Trial

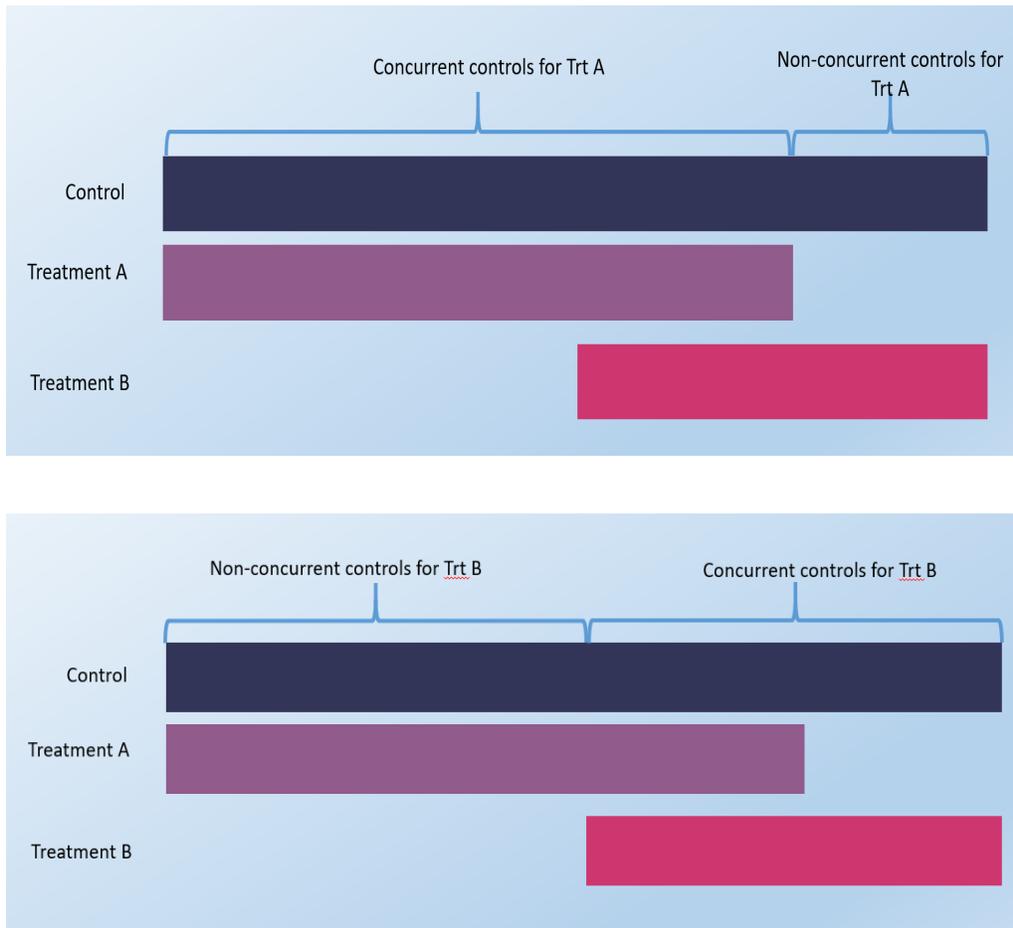

Apparently, it may be desirable to use both concurrent and non-concurrent control data to increase the efficiency of trial design and data analysis. However, statistical properties and assumptions on when and how non-concurrent control may be used are the main topic of discussion. In a previous open-forum discussion (October 8, 2020), the use of concurrent common control to compare with experimental treatments in a Master Protocol and its implications on Type I error rate were discussed (Sridhara et al. 2021b). The focus of this virtual discussion was to understand the statistical properties of data utilization from the non-concurrent control arm and its impact on Type I error rate in establishing treatment effect.

**Can we use non-concurrent common control to evaluate efficacy?**

*Academic Viewpoints by Panelists*

It is recognized that multi-arm trials with a common control can substantially increase the efficiency of drug development programs (Collignon et al. 2021). In rare diseases such designs can enable investigators to address questions that may not be otherwise feasible. Because of the staggered entry and exit of treatment arms and continued enrollment into the common control arm, the question arises whether the cumulative control data (non-concurrent and concurrent, or non-concurrent only) can be used in assessing the therapeutic effects of a treatment. The concept of using non-concurrent (NC) control data is similar to that of using historical control data. It is more likely to achieve efficiency gain with the use of NC control data from a randomized Master Protocol as opposed to the use of historical control since the NC control data has many standardized features of the randomized study and is collected within the same infrastructure. However, a main concern for the use of NC control data for assessing the effect of a treatment is that patients are randomized to the treatment and NC control at different times. A few potential confounding effects are of concern: prognostic factors may not be balanced, patients and investigators may not be blinded, and the lack of head-to-head randomization may introduce measured and unmeasured confounding effects that may be related to time trends. For example, internal (e.g., recruiting centers, inclusion/exclusion criteria, endpoint assessments) and external (e.g., standard of care, patient population) factors may introduce bias in data analysis when not modeled or accounted for.

Pooling of controls (NC and concurrent) can increase the Type I error rate, whereas, use of concurrent control only, ensures the control of Type I error even in the presence of time trends (the overall effects of confounders over time). Dynamic approaches, that pool data only if no substantial time trends are observed, have only minimal impact on Type I error if there is no or a very large time trend. However, some Type I error inflation is observed in intermediate cases (Jiao et al. 2019). One could consider modeling time drifts in the control group by including calendar time as a continuous or categorical covariate. These approaches control the Type I error rate if the model assumptions are correct (Lee and Wason, 2020). One needs also to pay attention to how much extrapolation of time and treatment effects is being done through the modeling approach because such methods rely on specific assumptions. Inclusion of NC control is a tradeoff between variance and risk of bias.

There is also a potential for information leakage due to completed treatment versus control comparisons regarding the effect of non-concurrent control which may lead to bias if it affects the decision to add new treatments to the Master Protocol. Validity of statistical inference relies on pre-specification of the analysis before data becomes available. If the protocol and the statistical analysis plan for new treatments are written based on such information, the principle of pre-specification is violated.

*Industry Viewpoints by Panelists*

Platform trials such as I-Spy 2, GBM-AGILE, and Precision Promise have used concurrent and NC controls to analyze data and make comparisons by modeling time-dependent confounders using Bayesian "time machine" (Saville and Berry, 2016), wherein a parameter is smoothed by normal dynamic linear model. Pooling of NC control data could increase bias. When time is considered as categorical unbiased estimation may be achieved. However, using "time machine" nearly unbiased estimates are obtained with inferential improvements (e.g., power, MSE) with a few caveats. For example, the NC controls must be randomized under the same protocol with the same inclusion/exclusion criteria, the same assessment visits, procedures, and data quality.

Therefore, the confounding effects between the concurrent and NC control can be captured by modeling time alone. Mathematical models such as Bayesian "time machine" use data from all arms in estimating the relative effects of any arm, can handle changing control arms, and enable response adaptive randomization to adjust for time. One of the benefits of the modeling time is that as the data grows, model gets smarter and more efficient and, therefore, the trial can enroll less patients to the control arm. Of course, such benefits assume that the time effect is sufficient for modeling measured and unmeasured confounders.

Another perspective was that when using NC control only, there are several risks that need to be considered: if the use is not justified, not able to know if concurrent control and NC control are similar or not, the risk of "cherry picking" while knowing the control performance in advance can introduce bias in treatment comparison. Use of NC control is similar to non-inferiority hypothesis testing where exchangeability is difficult to establish. Because the data are collected at a different calendar time, randomization does not ensure exchangeability of the distribution of prognostic factors between non-concurrent control and experimental arms.

Industry panelist also shared many of the concerns raised by academia, for example, that models are dependent on assumptions. Another practical comment was that if a sponsor wants to use all data for a submission, data-sharing can be challenging from a legal perspective.

*Regulatory Agencies' Viewpoints by Panelists*

Use of concurrent common control for evaluating treatment effect in Master Protocols are generally acceptable. However, a change in patient population, medical practice and a standard of care is of concern when using non-concurrent control. Typically, in trials sponsored by pharmaceutical companies, time trend is managed by the clinical trial design features such as blocked randomized designs and not by analysis. Non-concurrent control data information leakage could bias treatment effect estimate for treatments entering the protocol later. Understanding reason for time trend is important. Concurrent control data are needed, and the question is how available NC control data can be used appropriately as additional piece of information. Modeling time trend does not account for information leakage and other covariates that may confound the treatment effect. In the presence of potential time trend, comparison of the results of including and excluding NC control data in the analysis of the newly added interventions should be considered.

The move from concurrent control to NC control would be a big change in the regulatory decision making, bias in estimated treatment effect and a control of Type I error are issues to be considered. In rare diseases, NC control can provide good quality data and advantages of having a control arm in the trial compared to use of historical or other external control data. In evaluating treatments for rare diseases, one may be willing to consider less conservative Type I error control. However, regulators need compelling evidence that controls and patient population have not changed over time or need to understand why there is a change in a control arm.

**Highlights of Discussion**

The discussions in this virtual open forum were productive and covered different aspects of the use of concurrent and non-concurrent controls for platform trials. While non-concurrent control data are collected within the same framework as a concurrent control data and patients are randomized, because the data are collected at a different calendar time, randomization does not ensure balance in the distribution of both measured and unmeasured prognostic factors between non-concurrent control and experimental arms. Non-concurrent control and historical data share several sources of potential bias. Non-concurrent control may introduce bias due to different factors related to time, changes in standard of care, changes in patient population, changes in recruiting strategies, changes in assessment of endpoints, etc. Methods to address potential bias are available, however, they rely on specific assumptions and cannot control unobserved factors and, therefore, make analysis less reliable, particularly with potential information leaks based on interim decisions that reveal treatment effects for arms that are still under study. The choice of endpoints matter; for example, overall survival is a more objective endpoint, but to mitigate the immortal bias in time-to-event analysis might also be challenging. In more common indications and in a confirmatory setting, an acceptance of treatment claims based on a non-concurrent control is unlikely.

When historical data are used for comparisons in clinical trials, it is accepted that strict Type I error control is not possible. Similarly, in rare disease setting when it is impossible to run a large trial or in an exploratory setting, the use of non-concurrent control may potentially be considered on a case-by-case basis. For example, in a rare disease setting where the standard of care has not been changed for a long period of time and a time trend can be adequately modeled and with the understanding that the estimate of treatment effect may potentially be biased, use of non-concurrent control for treatment comparison in a platform trial may be acceptable.

This forum provided an opportunity to have open scientific discussions among diverse stakeholder group – academicians, international regulators, and pharmaceutical companies focused on emerging statistical issues in cancer drug development. We plan to continue with similar open forum discussions in the future on a variety of important topics that include statistical aspects in cancer drug development involving different stakeholders and a multi-disciplinary approach.

**Acknowledgement:** Authors thank Joan Todd (OCE FDA), Suman Sen (Novartis) and Jingjing Ye (Beigene) for taking the meeting minutes.

Martin Posch is a member of the EU Patient-centric clinical trial platform (EU-PEARL). EU-PEARL has received funding from the Innovative Medicines Initiative 2 Joint Undertaking under grant agreement Nº 853966. This Joint Undertaking receives support from the European Union's Horizon 2020 research and innovation programme and EFPIA and Children's Tumor Foundation, Global Alliance for TB Drug Development non-profit organisation, Springworks Therapeutics Inc. This publication reflects the authors' views. Neither IMI nor the European Union, EFPIA, or any Associated Partners are responsible for any use that may be made of the information contained herein.

Yuan Ji is a co-founder of Baysoft Inc., a company that provides consulting service to pharmaceutical and biotech companies. He is also an IDMC member for Astellas and consultant for Cytel. Yuan Ji had research contracts with Abbvie and Sanofi.

- Discussion". *Statistics in Biopharmaceutical Research,* DOI:10.1080/19466315.2021.1906743.

- Woodcock, J., and LaVange, L. M. (2017), "Master Protocols to Study Multiple Therapies, Multiple Diseases, or Both," *New England Journal of Medicine*, 377, pp. 62–70. DOI: 10.1056/NEJMra1510062.

# Appendix

## American Statistical Association Biopharmaceutical Section's

## Virtual Discussion on: Use of Non-concurrent Common Control for Treatment Comparisons in Master Protocols

Host: Statistical Methods in Oncology Scientific Working Group

December 10, 2020

8 am – 10 am EST (New York)

### Agenda

Meeting Moderators:

Dr. Qi Jiang, Seagen, Co-chair of ASA BIOP Statistical Methods in Oncology Scientific Working Group

Dr. Olga Marchenko, Bayer, Co-chair of ASA BIOP Statistical Methods in Oncology Scientific Working Group

Dr. Rajeshwari Sridhara, Oncology Center of Excellence, FDA

1. 8 am – 8:10 am: Welcome and Introduction
   - Dr. Olga Marchenko, Co-chair of ASA BIOP Statistical Methods in Oncology Scientific Working Group
   - Dr. Richard Pazdur and Dr. Rajeshwari Sridhara, Oncology Center of Excellence, FDA

2. 8:10 am – 8:50 am: Presentations
   - Prof. Martin Posch, Medical Statistics at the Medical University of Vienna
   - Dr. Scott Berry, Berry Consultants

3. 8:50 am – 9:50 am: Panel Discussion:

   Dr. Richard Pazdur (FDA), Dr. Marc Theoret (FDA), Dr. Yuan-Li Shen (FDA), Dr. Thomas Gwise (FDA), Dr. Kit Roes (EMA), Dr. Khadija Rantell (MHRA, UK), Mr. Andrew Raven (HC, Canada), Dr. Lorenzo Hess (Swissmedic); Dr. Richard Simon, Dr. Mary Redman (Fred Hutch), Dr. Yuan Ji (University of Chicago), and Dr. Cindy Lu (Biogen)

4. 9:50 am – 10:00 am: Concluding remarks
   - Dr. Qi Jiang, Seagen, Co-chair of ASA BIOP Statistical Methods in Oncology Scientific Working Group
   - Dr. Richard Pazdur and Dr. Rajeshwari Sridhara, Oncology Center of Excellence, FDA